# Cathodoluminescence of quantized energy states at AlGaN/GaN interface

Fatemeh Chahshouri[1,*], Masoud Taleb[1], Maximilian Black[1], Michael Mensing[2], Nahid Talebi[1,3,*]

[1]*Institute of Experimental and Applied Physics, Kiel University, 24098 Kiel, Germany*
[2]*Fraunhofer Institut für Siliziumtechnologies, ISIT, Fraunhoferstraße 1, 25524 Itzehoe, Deutschland*
[3]*Kiel, Nano, Surface, and Interface Science − KiNSIS, Kiel University, 24098 Kiel, Germany*
E-Mail: talebi@physik.uni-kiel.de; chahshouri@physik.uni-kiel.de

Abstract:

Recent progress in manufacturing high-electron-mobility transistors and optoelectronic devices highlights the necessity of understanding the charge dynamics and its impact on the optical properties inside heterostructures. Herein, we study the optical properties of $GaN/Al_{0.23}Ga_{0.77}N/GaN$ heterostructures using cathodoluminescence and photoluminescence spectroscopy. We explore the influence of generated secondary carriers after electron illumination and their penetration depth on the luminescence spectra. Our findings indicate that a higher laser power intensifies the photoluminescence response and establishes Fabry-Perot-like resonances. Furthermore, the intensity of the cathodoluminescence response shows a linear behavior versus the acceleration voltage and the current of electron beams for the yellow luminescence peak. A near-infrared cathodoluminescence peak (740 nm) is observed only when illuminating the sample with high currents that is attributed to the trapping of the secondary electrons within the Schottky barrier and the manipulation of the two-dimensional electron gas and the quantum-confined states within the barrier. Self-consistent Poisson-Schrödinger simulations verify this aspect. This research unveils the intricate charge dynamics associated with the interaction of electron beams with heterostructure systems, paving the way for innovative optoelectronic applications in semiconductor devices.

## Introduction

In the past few years, using semiconductors of group III-nitride materials and their junction has received attention in developing the next generation of power[1] and optoelectronic devices[2,3]. The exceptional properties of this semiconductor heterostructure (AlGaN/GaN), such as wide band gap and high saturation electron velocity[4], make GaN-based transistors[5] promising candidates for operating in low-resistance[6], high-frequency[7,8], high-power[9], and high-temperature[10]. Furthermore, the large conduction band offset[4] and strong polarization effect[11] at the interference of the semiconductors lead to the confinement of a high-density two-dimensional electron-gas[12] (2DEG) within the triangle quantum well[4]. This confinement enhances the electron mobility[13] to $2000\ cm^2 V^{-1} \cdot s^{-1}$ and electron density[13] to $3 \times 10^{13}\ cm^{-2}$ in the well, even without doping[11].

The broad luminescence response of the nitride materials in the near-infrared[14], visible[15,16], and ultraviolet[15] spectral ranges unleash their potential applications in fabricating light-emitting diodes[17,18] and semiconductor lasers[19]. Depending on the growth parameters and deposition methods, GaN thin films exhibit various defect-related luminescent[16] responses, including yellow[20,21], blue[20,21], and red[22] luminescence, alongside with direct band emission[15] luminesce at 3.4 eV. The yellow luminescence[23] (YL) impurity-related peak in the GaN, centered at 2.15 eV is the most common emission. This peak is associated with gallium vacancies ($V_{Ga}$) and arises from electron transition from a shallow donor to a deep acceptor level[16,23]. On the other hand, depending on the Al molar fraction, AlGaN exhibits a wider variety of point defects, such as nitrogen vacancy complexes[24] ($N_{Al}V_N$) and divacancy complexes[24] ($V_{Al}V_N$), compared to the GaN and AlN binary compounds. The other notable defect, which originates from the nitrogen vacancy complex ($V_N N_{Ga}$)[25], can result in zero-phonon line (ZPL)[25] emission at 1.46 eV (849 nm) and 1.04 eV (1170 nm) in the AlGaN.

Until now, defect-related and 2DEG luminescence have been widely studied using Photoluminescence[26,27] (PL) spectroscopy, where the density of photoexcited carriers are less than the equilibrium sheet carrier concentration[28]. Systematic studies under the influence of extra charge carriers[29,30] generated by high electron beam current together with cathodoluminescence (CL) spectroscopy at the interface of heterostructures, as shown here, will provide complementary information and allow for modulation of the sheet current density, and modification of the quantum levels at the Schottky barrier. In this study, we introduce CL as a method to explore the impact of the incident electrons on the induced electric field and the quantized energy levels in the 2DEG region. Both CL and PL spectroscopy techniques are employed to analyze the structural and optical properties of undoped AlGaN/GaN heterojunctions. The study investigates the effects of laser power, electron-beam energy, and electron current to gain insights into carrier dynamics and optical emissions from defects and the 2DEG in the heterostructure. Monte-Carlo simulations[31] are employed to elucidate the depth-resolved electron trajectory and concentration of absorbed electrons inside the heterostructure. Finally, the energy band diagram and eigenenergy states under the influence of extra charge carriers are determined using the self-consistent Poisson-Schrödinger calculation[32].

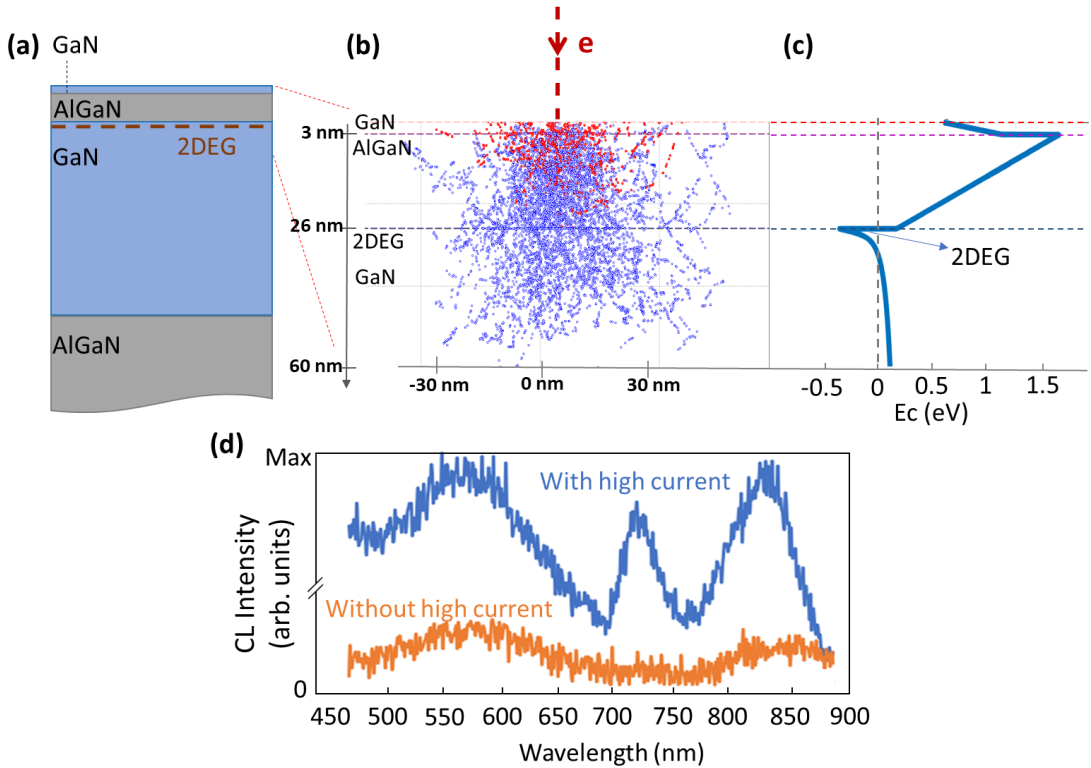

**Fig. 1.** Schematic diagram of the fabricated AlGaN/ GaN device and calculated conduction band. (a) The MOCVD growth sequence starts with a 3.5 μm thick AlGaN layer on a Si substrate, followed by a 350 nm thick GaN buffer, a 23 nm AlGaN barrier, and a 3 nm thin GaN cap layer. (b) Electron trajectories with segments of elastic- and inelastic-scattered secondary electrons (blue dots) and backscattered electrons (red dots), excited by a 2 kV electron beam. (c)

The calculated potential profile is associated with the conduction band in the upper layers of the heterostructure. (d) The monochromatic cathodoluminescence (CL) maps is taken with 20 kV acceleration voltage at low and high currents (low current: 4 nA; high current: 14 nA)

The multilayer sample studied here was grown by metal-organic chemical vapor deposition (MOCVD) on an n-doped (111) Silicon substrate. The heterostructure comprises a non-conductive 3.5 μm Carbon-doped AlGaN layer. Subsequently, a 350 nm thick unintentionally doped GaN buffer layer was deposited, followed by an $Al_{0.23}Ga_{0.77}N$ layer spacer with a thickness of 23 nm. Finally, a 3 nm passivate cap layer of GaN was deposited (Fig. 1(a)).

Electrons illuminating the sample lead to electron-electron collisions and electron-hole pairs generation within the semiconductor. Monte-Carlo[31] simulation is used to study the inelastic scattering events inside the material. The trajectory of electrons in the upper layers of the junction is depicted in Fig. 1(b). The penetrating electrons inside the layers undergo both elastic and inelastic interactions, resulting in lateral and depth expansion, as well as the generation of secondary electrons. The maximum penetration depth at an excitation energy of 2 keV is approximately 60 nm, with the maximum electron energy loss occurring at about one-third of this value (close to the 2DEG region). To further understand the subtle influence of each layer thickness on the emission process, we investigate the energy band diagram and the charge distribution inside the structure. Here, the self-consistent solution of Poisson and Schrodinger equations[32] is utilized to calculate the energy profile in the $GaN/Al_{0.23}Ga_{0.77}N/GaN$ structure. The spatio-energy distribution of the conduction band of this heterostructure near the interface is illustrated in Fig. 1(c). Owing to the discontinuity and conduction band offset between the AlGaN and GaN layer, a triangular quantum well is formed at the hetero-interface. Consequently, electrons become confined and are separated from the positively charged donors by a potential barrier.

Spatially resolved scanning electron microscope cathodoluminescence (SEM-CL) experiments were conducted using a Delmic Instruments CL[33–37] system installed on a ZEISS scanning electron microscope. A room-temperature CL spectrum was obtained with a beam energy of 20 keV and a beam current of 4 nA and 14 nA (Fig. 1. d). The spectrum of low current electrons exhibits GaN YL emission at 575 nm, and the AlGaN ZPL centered at approximately 850 nm. By increasing the electron beam current, a luminance peak at 740 nm becomes apparent. Since this luminance significantly depends on the number of electrons, and it appears just in high current, we attribute this to recombination related to electrons captured in the 2DEG region. To gain further insight into the role of electron concentration and density in the luminescence response, especially at 740 nm, we perform additional CL measurements and simulations as shown below.

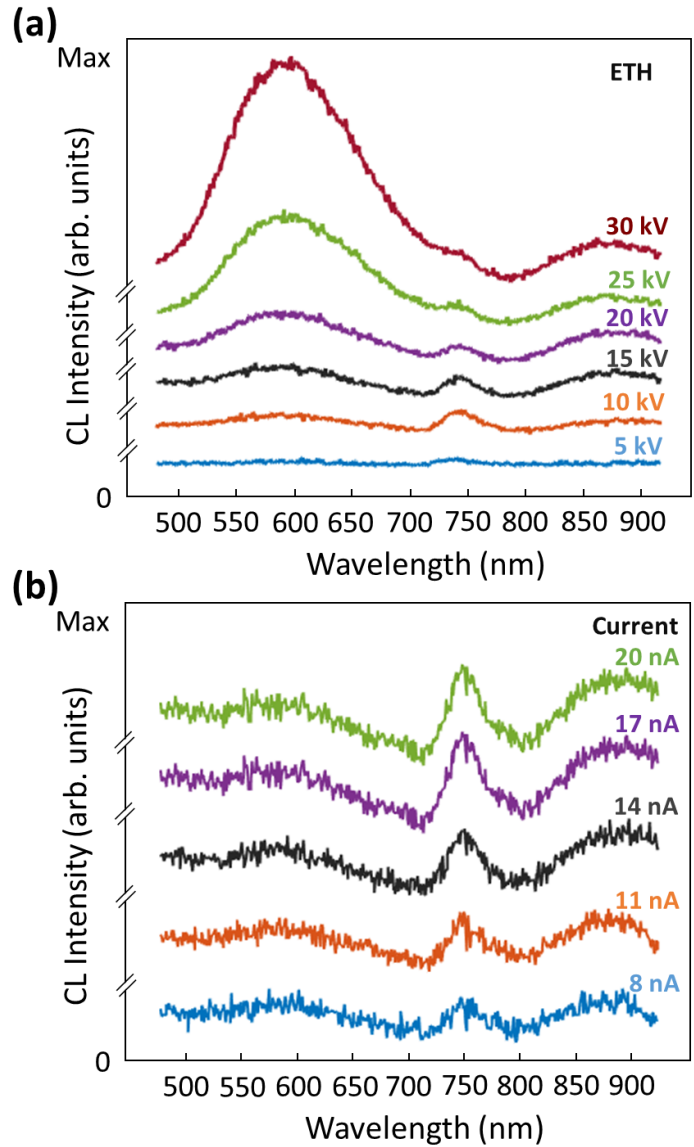

**Fig. 2.** Room temperature CL spectra of a heterostructure. (a) CL spectrum obtained at depicted electron beam acceleration voltages with a beam current of 12 nA. (b) Acquired CL spectrum corresponding to the 10 kV acceleration voltage at different depicted beam currents.

The supplementary CL measurements were performed in two ways: first, by varying the electron-beam energy while keeping the current constant at 12 nA beam current, and second, by modifying the electron beam current at a fixed acceleration voltage (10kV). Analyzing the CL spectroscopy data of the heterostructures at different electron beam voltages ranging from 5 to 30 kV allows us to explore the properties of defects and 2DEG at the interface. Under this irradiation condition, with high current, gradual increase in the intensity of the defect-based luminescence (YL and ZPL) is observed by increasing the acceleration voltage (as shown in Fig. 2(a)). Since these defects are localized state[38], a higher beam power enables electrons to penetrate more profoundly into the sample and occupying more impurity states[39]; hence, causing a stronger luminescence response[40].

Conversely, the middle luminescence intensity (at 740 nm) shows a non-linear response to the acceleration voltage. Increasing the voltage from 5 to 15 kV enhances this intermediate luminance intensity. However, at voltages exceeding 15 kV, the YL defect-based luminescence peak is the prominent response. The results show that the intensity of the mentioned luminescence directly correlates with the number of carriers inside the 2DEG region. By increasing the beam energies without changing its current, the kinetic energy of the electrons increases, allowing them to penetrate more deeply into these layers. Consequently, the trapping of secondary electrons in the 2DEG region decreases as the voltage increases. This leads to decrease in the luminescence spectral intensity (The Monto Carlo simulation in the next session shows the electron trajectory inside the sample).

Emphasizing the role of sheet current density, increasing the electron current at an initial energy of 10 keV significantly amplifies the trapped charge particles within the Schottky barrier. The increase in the current indicates an increase in the number of impinging electrons and consequently, the secondary electrons. This leads to a higher concentration of carriers in the quantum well region. As a result, more carriers are available to recombine with the trap states, therefore sharper luminescence peaks are observed (Fig. 2(b)). Using the CL spectroscopy, we have demonstrated that the intensities of the luminescent bands depend not only on the defect concentration but also on the electron-hole pair density and injection rate[41].

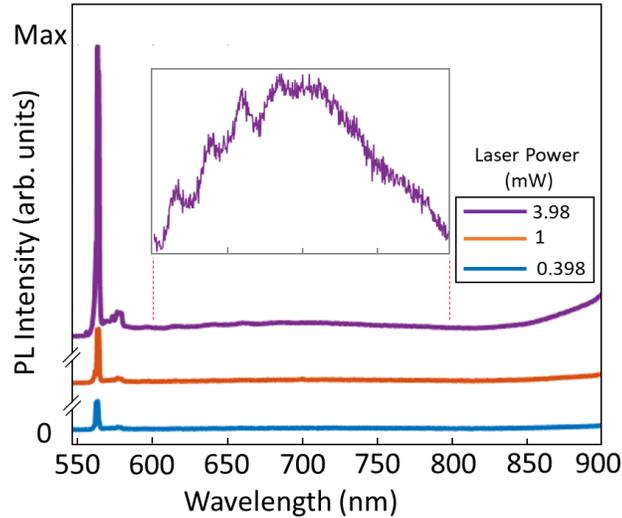

**Fig. 3.** Power-dependent photoluminescence (PL) spectra of a sample recorded at 2 min acquisition time. (a) 0.398 (mW), (b) 1(mW), and (c) 3.98 (mW) laser intensities. The inset shows the PL spectra of the sample excited at the highest intensity within the zoomed wavelength range, where ripples exhibit a Fabry-Perot-like resonance, due to the reflections from the boundaries of the structure.

For the PL measurements, the sample was excited by a continuous-wave laser emitting at the wavelength of 532 nm. Fig. 3 shows the PL spectra of the sample. Similar to the CL results, two common defect-related luminescence features are observed, and the absolute PL intensities strongly correlate with the density of the generated electron-hole pairs[40]. Increasing the laser power results in the generation of more carriers, leading to the occurring of more electron-hole recombination events[40]. The ripples observed at the shoulder of the YL for the intense PL spectra suggest the occurrence of the total internal refection of the generated luminescence at the interface that sets the microcavity effect[42]. Furthermore, within the spectral range of 600 nm-700 nm, a broad spectral feature and multiple weak ripples are observed that highlight the influence of coulomb scattering centers and the Bremsstrahlung radiation[43]. However, unlike CL measurement, the PL spectroscopy does not exhibit a sharp response at 740 nm. This happens because electron-beam excitation typically results in carrier generation rates orders of magnitude higher than those from typical laser excitation, that can lead to the superbunching effect[30,44]. The density of photoexcited carriers is significantly lower than the equilibrium electron concentration, whereas electron excitation changes the transient carrier density significantly, due to the generation of secondary electrons with arbitrary kinetic energies.

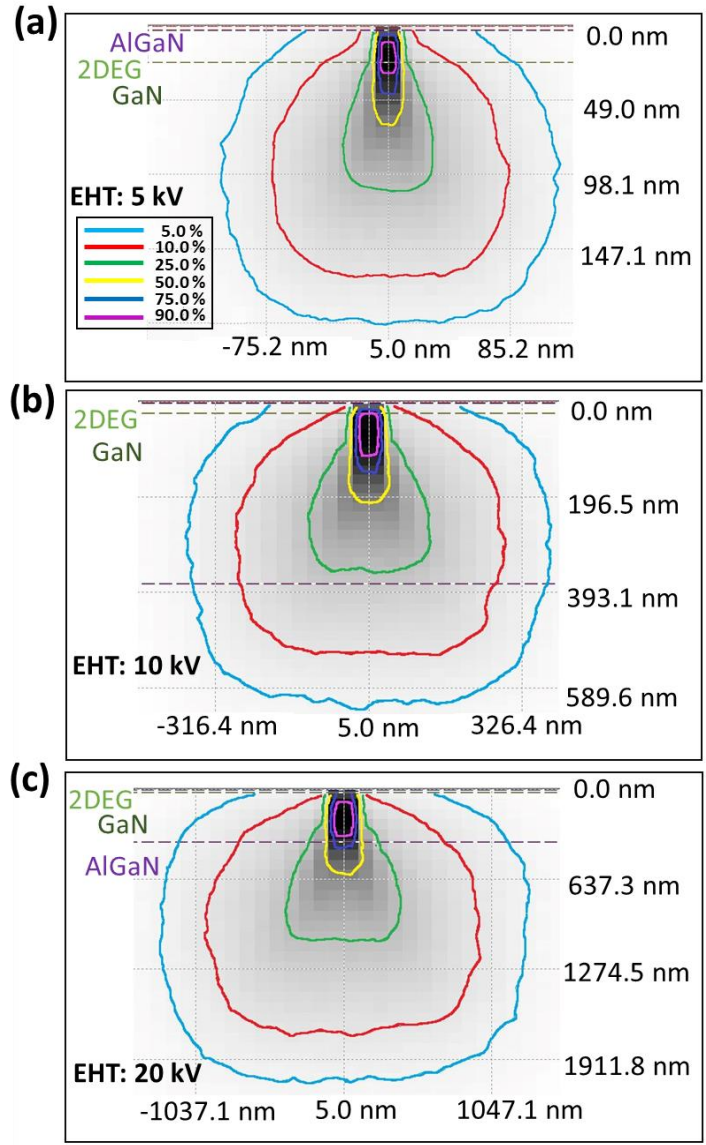

**Fig. 4.** Monte Carlo simulations of the spatial distribution of the absorbed energy in the heterostructure. Electron incident kinetic energy is (a) 5 keV, (b) 10 keV, and (c) 30 keV. The number of incident electrons is 100000, and a gray shading demonstrates the density of the absorbed energy in each region, with dark and light shadows meaning regions with maximum and minimum energy losses, respectively. The cross-section contour energy lines show the spatial profile of the inelastically scattered electrons absorbing the depicted percentages of the initial kinetic energy.

To further investigate the properties of the 2DEG and defect-related emissions, we calculated a depth-resolved electron energy loss map inside the heterojunctions. The energy dissipation of the incident electrons is simulated with CASINO Monte Carlo software[31], which includes the simulation of 2D samples. The Monte Carlo method involves tracing the path of the electrons within the sample. This package simulates elastic scattering events and approximates inelastic events[45] through the mean energy loss between two elastic scattering interactions[46]. In the simulation process, we initially added multilayers with thicknesses specified in the fabrication method. Subsequently, we defined parameters for the electron beam in the simulation domain, including beam diameter (10 nm), polar angle of the incident beam (vertical electron beam illumination), initial electron kinetic energy, and the number of incident electrons). Finally, we employed the Mott approximation[47] to interpolate the total cross section and the Joy–Luo method to account for ionization potential, collision, and inelastic energy loss[45].

Electrons arriving to the sample start penetrating into it. Depending on its initial energy and the type of the material, it experiences energy dissipation and scattering. Consequently, it undergoes lateral and vertical diffusion[48]. As demonstrated in Fig. 4, the CASINO simulations reveal a significant electrons energy loss as they propagate inside the heterostructure. Contour energy lines represent the amount of the absorbed electrons energy in the sample; darker gray shading signifies higher electron density, which decreases spatially to lighter shading, that indicates lower density. Fig. 4(a) illustrates the energy distribution inside the sample interacting with 5kV electron beams. The purple energy line demonstrates intense absorption and concentration of electrons at the AlGaN/GaN interface, accounting for 90% of their initial energy. This heightened concentration of electrons results in an increased number of carriers in the 2DEG region and enhances luminescence peak at 740 nm (as observed in Fig. 2(a) the only prominent luminesce is occurred at this wavelength). After illuminating the sample with 10kV electron beams (Fig. 4(b)), electron density intensifies mostly in the GaN layer, and it reaches AlGaN as well. In this condition, the substantial concentration of electrons (90% counter-energy line) remains at the border of the AlGaN/GaN interface; however, most of the electrons diffuse deeper into the interior area of the GaN and AlGaN. Consequently, it causes a notable rise in the luminescence intensity from both YL and 2DEG emission (as demonstrated in Fig. 2(a)). Moreover, the ZPL luminesce at 880 nm originating from the AlGaN layer is weakly observed. Applying 20 kV (Fig. 4(c)) voltages, high energy electrons traverse the GaN and reach the second AlGaN layer (75% counter-energy line placed in the second AlGaN layer). Fast transient of the high energy electrons from the interface leads to a reduction in the density of the absorbed electrons in the Schottky barrier that hosts 2DEG; thus, there is a less pronounced intermediate luminescence at 740 nm and enhanced luminescence from both GaN and AlGaN defect center (as shown in Fig. 2(a)).

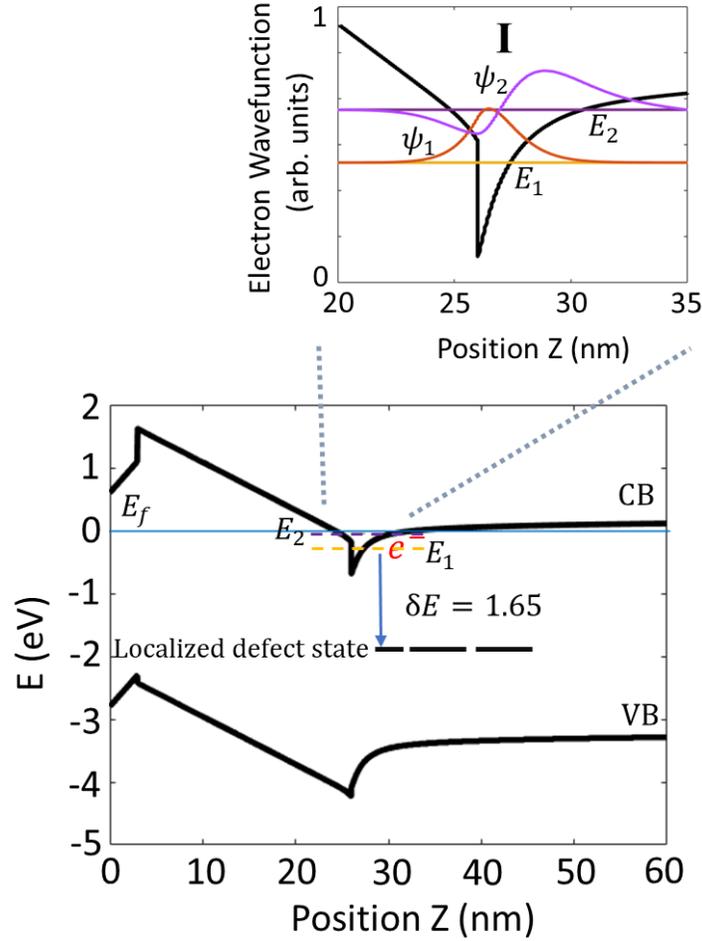

**Fig. 5.** The calculated band structure and quantum wave function obtained using a self-consistent Poisson-Schrödinger simulation at the GaN/AlGaN/GaN heterojunctions (see text for details). Two sub-band electronic levels occupied under the influence of $2 \times 10^{13} (\text{cm}^{-2})$ sheet carrier density. (I) The enlarged conduction band profile in the 2DEG region, and the calculated wave function for the first and second quantum states in the triangular-shaped quantum well. The symbols CB, VB, and $E_f$ represent conduction band, valence band, and Fermi level energy, respectively.

To better understand the physical properties behind the extraordinarily and sharp cathodoluminescence peak, we employed a self-consistent Poisson-Schrödinger solver to calculate the band diagram, carrier concentrations, and electron wave function in the heterostructure[32]. The thicknesses of layers[38], and the Al mole fraction[38] in the barrier layer are taken from the experimental values, then the band energy profile and electron wave function are obtained by solving the one-dimensional system using the finite difference method with nonuniform mesh where both spontaneous and piezoelectric polarization[11] are considered. The calculated equilibrium sheet carrier density inside the triangle quantum well for the mentioned AlGaN/GaN heterostructure, without applying voltage, is $0.9 \times 10^{13} (\text{cm}^{-2})$. When exposed to electron illumination, the concentration of extra electrons[30] in the 2DEG area and bunching effect leads to increasing in the equilibrium sheet density and the image potential, hence altering the shape of the triangular quantum well. As shown in Fig. 5, considering the influence of higher sheet density as $2 \times 10^{13} (\text{cm}^{-2})$, the quantized energy level is shifted toward the potential notch. This shift results in a higher probability of electron-hole recombination between the ground energy and the localized defect state. Therefore, this distribution allows electron transitions to the localized defects[39] center in the GaN. Fig. 5 illustrates the two bound wave functions. The intensity of the CL peak is correlated with the overlap between the wave functions of the localized electronic states associated with the trapped states in the Schottky barrier and the atomic structure. In the ground electronic state, localization of wave function occurs within the quantum well, with a non-uniform weak tail distribution extending towards the outer edge of the AlGaN/GaN interface. Our results show that even at high densities as high as $2 \times 10^{13} (\text{cm}^{-2})$, we can still expect nearly good confinement of electrons in the quantum well (occupation of two sub-bands). This is an essential point because the quick occupation of the sub-bands results in a loss of actual two-dimensional behavior of the gas and a lowering of the mobility due to inter-sub-band scattering.

In summary, in this work, we study the luminescence from undoped GaN/AlGaN heterostructure. The PL and CL measurements on the sample unravel the defect-based luminescence (YL and ZPL) after laser and electron excitation. Furthermore, electron excitation indicates another luminescence response from 2DEG emission. The highly concentrated electrons at the interface of AlGaN/GaN thin films increase the carrier sheet density in the 2DEG region. Reshaping the band profile and changing the position of the energy levels in the GaN buffer quantum well, intensifies the probability of electron transition to the deep acceptor point defects. Our findings show that strengthening the luminescence by increasing the charge carriers can be achieved by controlling the excitation energy to have maximum absorption in the 2DEG region and applying a higher current. The results are supported by the excellent correlation between the experimental luminesce measurement with the electron trajectory and the energy band diagram calculations inside the heterostructure. Our work provides insight into the role of the charge dynamics and spatial diffusion in identifying the defect, and the 2DEG-based emission for controllable quantum emission generation, and metastable state shelving for fabricating the next generation of light-emitting diodes.

## Acknowledgement:

This project has received funding from the European Research Council (ERC) under the European Union's Horizon 2020 research and innovation program under grant agreement no. 802130 (Kiel, NanoBeam) and grant agreement no. 101017720 (EBEAM).